\documentclass[]{aa}

\usepackage{amsmath}
\usepackage[dvipsnames]{xcolor}
\usepackage{hyperref}

\providecommand{\dif}{\mathrm{d}}
\providecommand{\OmegaK}{\Omega_{\mathrm{K}}}
\providecommand{\rg}{r_{\mathrm{G}}}
\providecommand{\qpm}{\tilde{q}}
\providecommand{\RP}{\mathrm{RP}}
\providecommand{\LT}{\mathrm{LT}}

\hypersetup{colorlinks=true, allcolors=blue}

\newcommand{\orcid}[1]{\href{https://orcid.org/#1}{\includegraphics[width=8pt]{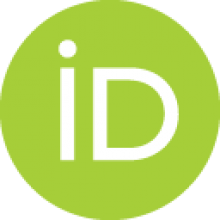}}}

\begin{document}

\title{Lense-Thirring precession of neutron-star accretion flows:\\ 
 Relativistic versus classical precession}
\titlerunning{Lense-Thirring precession of neutron-star accretion flows}

\author{
	Gabriel T\"{o}r\"{o}k\inst{1}\orcid{0000-0003-3958-9441}
	\and
    Martin Urbanec\inst{1}\orcid{0000-0001-9635-5495}
    \and
    Monika Matuszkov\'a\inst{1}\orcid{0000-0002-4193-653X}
    \and
    Gabriela Urbancov\'a\inst{1}\orcid{0000-0002-4480-5914}
    \and
    Kate\v{r}ina Klimovi\v{c}ov\'a\inst{1}\orcid{0000-0002-0930-0961}
    \and
    Debora Lan\v{c}ov\'a\inst{1,2}\orcid{0000-0003-0826-9787}
    \and 
    Eva \v{S}r\'{a}mkov\'{a}\inst{1}\orcid{0009-0000-7736-6180}
    \and
    Ji\v{r}\'{\i} Hor\'{a}k\inst{3}\orcid{0000-0002-7635-4839}
	}

\institute{
    Research Centre for Computational Physics and Data Processing, Institute of Physics, Silesian University in Opava, Bezru\v{c}ovo n\'am.~13, CZ-746\,01 Opava, Czech Republic
    \and
    Nicolaus Copernicus Astronomical Center of the Polish Academy of Sciences, Bartycka 18, 00-716 Warsaw, Poland
    \and
    Astronomical Institute of the Czech Academy of Sciences, Bo\v{c}n\'{\i} II 1401, CZ-14100 Prague, Czech Republic
}
          
\authorrunning{T\"{o}r\"{o}k et al.}

\abstract{
The vertical (Lense-Thirring) precession of the innermost accretion flows has been discussed as a sensitive indicator of the rotational properties of neutron stars (NSs) and their equation of state because it vanishes for a non-rotating star. In this work, we apply the Hartle-Thorne spacetimes to study the frequencies of the precession for both geodesic and non-geodesic (fluid) flows. We build on previous findings on the effect of the NS quadrupole moment, which revealed the importance of the interplay between the relativistic and classical precession. Because of this interplay, the widely used Lense-Thirring metric, linear in the NS angular momentum, is insufficient to calculate the behaviour of the precession frequency across an astrophysically relevant range of NS angular momentum values. We find that even for a moderately oblate NSs, the dependencies of the precession frequency on the NS angular momentum at radii within the innermost accretion region have maxima that occur at relatively low values of the NS angular momentum. We conclude that very different groups of accreting NSs --- slow and fast rotators --- can display the same precession frequencies. This may explain the lack of evidence for a correlation between the frequencies of the observed low-frequency quasiperiodic oscillations and the NS spin. In our work, we provide a full, general description of precession behaviour, and also examples that assume specific NS and quark star (MIT bag) equation of state. Our calculations are reproducible using the associated Wolfram Mathematica notebook.
}


\maketitle

\section{Introduction} \label{sec:intro}

Neutron star low-mass X-ray binaries (NS LMXBs), which consist of an NS accreting matter from a less massive companion star, are invaluable for studying extreme environments and fundamental physics of dense matter \citep{klis2006,lewin2010}. Among the various phenomena observed in these systems, the X-ray variability exhibits a rich and diverse phenomenology.

On short timescales, NS LMXBs often display a complex of quasi-periodic oscillations (QPOs) in the low- and high-frequency (LF and HF) bands. These features, represented by multiple peaks in the Fourier power-density spectra,  span from a few hertz to a few tens of hertz for LF QPOs and from a few hundred hertz up to 1.3 kHz for HF QPOs. The HF QPOs often consist of two peaks, one of which is sometimes highly coherent, with quality factors (the height of the peak at its half-width) reaching up to $Q\sim300$, and high root mean square amplitude reaching up to $30\%$ \citep[][]{2005MNRAS.361..855B,barret2006,barret2012}. The peaks within the LF QPO complex have relatively low coherence but still notable amplitudes \citep{barret2005, mendez2006, barret2012}. Similar behaviour is observed in the black hole LMXBs. However, the  QPOs are weaker there, and the HF QPOs represent somewhat elusive features with small amplitudes and coherences. Commonly to both classes of sources, HF and LF QPO frequencies seem to be well correlated \citep{stella1999,2002ApJ...580..423M,klis2006,2006ARA&A..44...49R,2016ASSL..440...61B}.

It is frequently assumed that the QPOs emerge from the orbital motion in the vicinity of the central NS \citep[][]{barret2006,klis2006}. A large variety of orbital models of QPOs have been proposed assuming either local effects within the accretion disc or the global motion of the disc itself \citep[e.g.,][]{1990ApJ...358..538K,1999ApJ...526..953K,2008A&A...487..527C,kato2001, wagoner2001}.  Many of them deal with an inner thick accretion flow (torus)  and its epicyclic oscillations \citep{abramowicz2001, kluzniak2001}, or other modes of torus oscillations \citep{2003MNRAS.344L..37R,rezzolla2003, blaes2006, fragile2016,ingram2010,2017MNRAS.467.4036M,2018MNRAS.474.3967D}.

At the end of the millennium, the relativistic precession (RP) model was proposed \citep{1998ApJ...492L..59S, stella1999,1999ApJ...524L..63S}, which relates the HF and LF QPOs to the radial (periastron) and vertical (Lense-Thirring, LT) precession of inhomogeneities orbiting within the innermost accretion flow. They noticed that the vertical precession emerges as a critical tool for probing the characteristics and internal structure of NSs. Notably, the precession driven by the relativistic frame-dragging effects of the NS's rotation should be absent in non-rotating NSs, making it possibly a distinctive indicator of NS rotation and equation of state (EoS). The concept of the RP model was later discussed in many works, including \citet{ingram2009, ingram2010, ingram2011, middleton2019, fragile2020}, and \citet{bollimpalli2024} where the authors focused on the vertical precession along with related results of numerical simulations of oscillations of accretion flows and their comparison with observations of LF QPOs. 

The predictions of the RP model and models predicting similar observable frequencies have been extensively studied in relation to both HF and LF QPOs in a variety of astrophysical sources \citep[][]{1998ApJ...492L..59S,stella1999,2012MNRAS.427..595M,2014MNRAS.437.2554M,2022MNRAS.517.1469M,2016ApJ...825...13S,2016A&A...586A.130S,2016MNRAS.457L..19T,torok2016,2018MNRAS.473L.136T,2018MNRAS.479..426V,2020MNRAS.496.5262V,2020A&A...643A..31K,2020ApJ...899..139M}. While in the former case the expected frequencies in NS sources seem to agree with the observations \citep[][]{2007MNRAS.376.1133B,2011ApJ...726...74L,torok2022}, in the latter case a lack of strong correlation between the frequencies of the LF QPOs and the NS spin was found \citep[][]{2012ApJ...759L..20A,2017MNRAS.465.3581V,2019MNRAS.490.5270V,2021MNRAS.502.5472D}.

In our previous works, we studied the oscillations of inner fluid accretion flow using an analytic approach, focusing on the radial precession and its possible relation to the HF QPOs \citep{2016MNRAS.457L..19T, torok2022}. In the recent papers of \cite{mat2024b,mat2024a}, we further provided tools to investigate the vertical precession. Here, we follow our most recent work and investigate the behaviour of the vertical precession frequencies.

\section{Approach to rotating oblate compact stars}

In the context of the RP model, the vertical epicyclic frequency and the frequency of LT precession of the geodesic motion around rotating oblate NSs were investigated by \cite{1999ApJ...513..827M}, who calculated the associated post-Newtonian formulae, studied numerical models of rotating NSs, and compared the model predictions with the data. The behaviour of the precession frequencies and possible observable signatures of the NS structure were later investigated by \cite{2012MNRAS.422.2581P}, who focused on the effects of multipole moments of spacetime. The study emphasised that, even for realistic NS models, the dependence of the precession frequency on the rotational frequency of the NS can be non-monotonic, and that the vertical epicyclic frequency can be greater than the Keplerian \citep[see also][]{2012PhRvL.108w1104P,2015MNRAS.454.4066P}.

The complex physics behind the behaviour of the geodesic epicyclic frequencies around rotating oblate compact stars was further explored in the works of \cite{kluzniak2013,kluzniak2014}, and \cite{2014PhRvD..89j4001G}, who , motivated by the works of \cite{2001PhRvD..63h7501Z} and \cite{2002A&A...381L..21A}, calculated analytical formulae describing the epicyclic frequencies of rotating oblate bodies in Newtonian gravity, namely the Maclaurin spheroids, and compared them with those calculated for rotating quark stars (QSs) using a numerical general-relativistic approach. They showed the importance of the interplay between the relativistic effects and the effects arising purely from the spatial distribution of the gravitating mass. In terms of the LT precession, this interplay manifests itself as a competition between the relativistic and classical precession. The results of these works were consequently extended in some aspects by \cite{tsang2016}, who focused on the self-trapping of discoseismic corrugation modes in NS spacetimes, and \cite{2022MNRAS.515.6164B}, who calculated epicyclic frequencies in the case of Newtonian stars composed of concentric spheroids.

Here, we generalised the previous results. Using a fully analytic, general relativistic approach, we examined LT precession in relation to any rotating, oblate compact stars, including NSs and QSs. Furthermore, we investigated not only frequencies associated with geodesic motion, but also those associated with non-geodesic fluid flows.

\subsection{Spacetime}

We study spacetime geometry around a rotating, axisymmetric compact object governed by general relativity in terms of the Schwarzschild coordinates $\{t, r, \theta, \varphi\}$. The corresponding metric tensor captures the curvature induced by the star's mass and rotation, and is represented by\footnote{In the adopted framework, the units of measurement are chosen so that both the speed of light $c$ and the gravitational constant $G$ are equal to one ($c = G = 1$). Distances are measured in terms of the gravitational radius, $\rg = GM/c^2$, and the metric signature used is $(-,+,+,+)$.}:

\begin{equation}
    \dif s^2 = g_{tt} \dif t^2 + 2 g_{t\varphi} \dif t \dif \varphi + g_{rr} \dif r^2 + g_{\theta\theta} \dif \theta^2 + g_{\varphi\varphi} \dif \varphi^2.
\end{equation}

To handle the calculation on the NS spacetime background, the LT metric \cite[][]{1918PhyZ...19..156L} is often used. However, this metric is only linear in the NS angular momentum. Here we adopt the Hartle-Thorne metric \citep[HT, ][]{hartle1967, hartle1968}, which takes into account the star’s mass $M$, angular momentum $J$, and quadrupole moment $Q$. To simplify the expressions, we use the dimensionless angular momentum and quadrupole moment, $j = J/M^2$ and $q = Q/M^3$. 
The components of the HT metric in notation adopted here \citep{abramowicz20003,urbancova2019} are, along with the components of the linear metric, given in a Wolfram Mathematica notebook associated with this work.\footnote{\url{https://github.com/Astrocomp/LTprecessionHT}}

\subsection{Quadrupole parameter}

In principle, a non-rotating NS can have non-zero oblateness. Here, we assume the oblateness to be entirely rotationally induced, and thus to vanish when the angular momentum of the star reaches zero. 
To study the effects of the oblateness and angular momentum on the orbital motion, we utilise the quadrupole parameter,
\begin{equation}
\qpm\equiv q/j^2. 
\end{equation}
This quantity is determined by the given NS EoS and $M$ and does not depend on the NS rotational frequency \citep[e.g.,][]{urbancova2019}. Furthermore, in terms of the spacetime geometry, its value determines whether the influence of the $g_{t\varphi}$ component of the metric tensor and the associated frame-dragging effects, which are linear in $j$, dominate or can be overtaken by the other components. This occurs as $j$ increases, and as the effects given purely by the spatial distribution of the gravitating mass associated with $j^2$ become important.

\subsection{Range of spacetime parameters}

The range of parameters considered here is constrained by the observed properties of neutron stars in accreting systems, with masses ranging from $1.4M_\odot$ to $2.5M_\odot$ \citep{strohmayer1996, lattimer2001} and rotational frequencies corresponding to millisecond pulsars. Previous studies of NS models have shown that the maximum dimensionless angular momentum is approximately $j_{\text{max}} \approx 0.7$, with a corresponding quadrupole parameter $\qpm$ ranging from $\qpm~\sim~1.5$ for a very massive (compact) NS to $\qpm~\sim~10$ for a low-mass NS depending on the NS equation of state  \citep[][]{lo2011,urbancova2019}.

\section{Orbital motion}

Properties of the orbital motion of a free test particle provide basic insights into gravitational interactions and the behaviour of accretion flows near compact objects.\footnote{Although our initial focus in this section is on geodesic motion, we note up front that the complete set of formulas presented in the associated Mathematica notebook is generally applicable to a non-geodesic fluid motion in thick accretion tori \citep{mat2024b}.  In the limit of zero torus thickness, this set reduces to formulae describing geodesic fluid motion or test particle motion. The geodesic case was considered in \cite{urbancova2019}. In their paper, which benefited from the introduction of $\tilde{q}$, the geodesic formulae originally derived by \cite{abramowicz20003} were presented in a compact, physically self-consistent notation.} In the simple case of  Schwarzschild spacetime, which is spherically symmetric and static, two fundamental orbital frequencies corresponding to a radially perturbed circular orbital motion exist:

 The azimuthal (Keplerian) angular frequency $\OmegaK$, and the radial epicyclic angular frequency $\Omega_r$,
\begin{equation}
    \OmegaK = \sqrt{\frac{M}{r^{3}}},
\quad
    \Omega_r^2 = \frac{M}{r^3}\left( 1- \frac{6M}{r} \right).
    \end{equation}
Accordingly, the relativistic precession arises, which manifests as a gradual advance in the orbit’s periastron, with angular frequency
\begin{equation}
\Omega_{\RP}^0 = \OmegaK - \Omega_r.
\end{equation}

In the HT spacetimes, the azimuthal and radial frequencies are then modified due to the additional frame-dragging and quadrupole contributions associated with the slow NS rotation. The azimuthal frequency reads \citep{abramowicz20003}
\begin{equation}
    \OmegaK = \frac{M^{1/2}}{r^{3/2}} \left[ 1 - j \frac{M^{3/2}}{r^{3/2}} + j^2 \alpha_1 (r) + q \alpha_2 (r) \right],
\end{equation}
and the radial frequency reads \citep{abramowicz20003}
\begin{equation}
     \Omega_r^2 = \frac{M}{r^3}\left( 1- \frac{6M}{r} \right) \left[ 1 + j \gamma_1(r) - j^2 \gamma_2(r) - q \gamma_3(r) \right].
\end{equation}
 The above terms $\alpha_1(r)$, $\alpha_2(r)$, $\gamma_1(r)$, $\gamma_2(r)$, and $\gamma_3(r)$ reflect the effects induced by the NS rotation and are provided in their complete form in the associated Wolfram Mathematica notebook.

\begin{figure*}[t]
    \centering
    \includegraphics[width=0.97\linewidth]{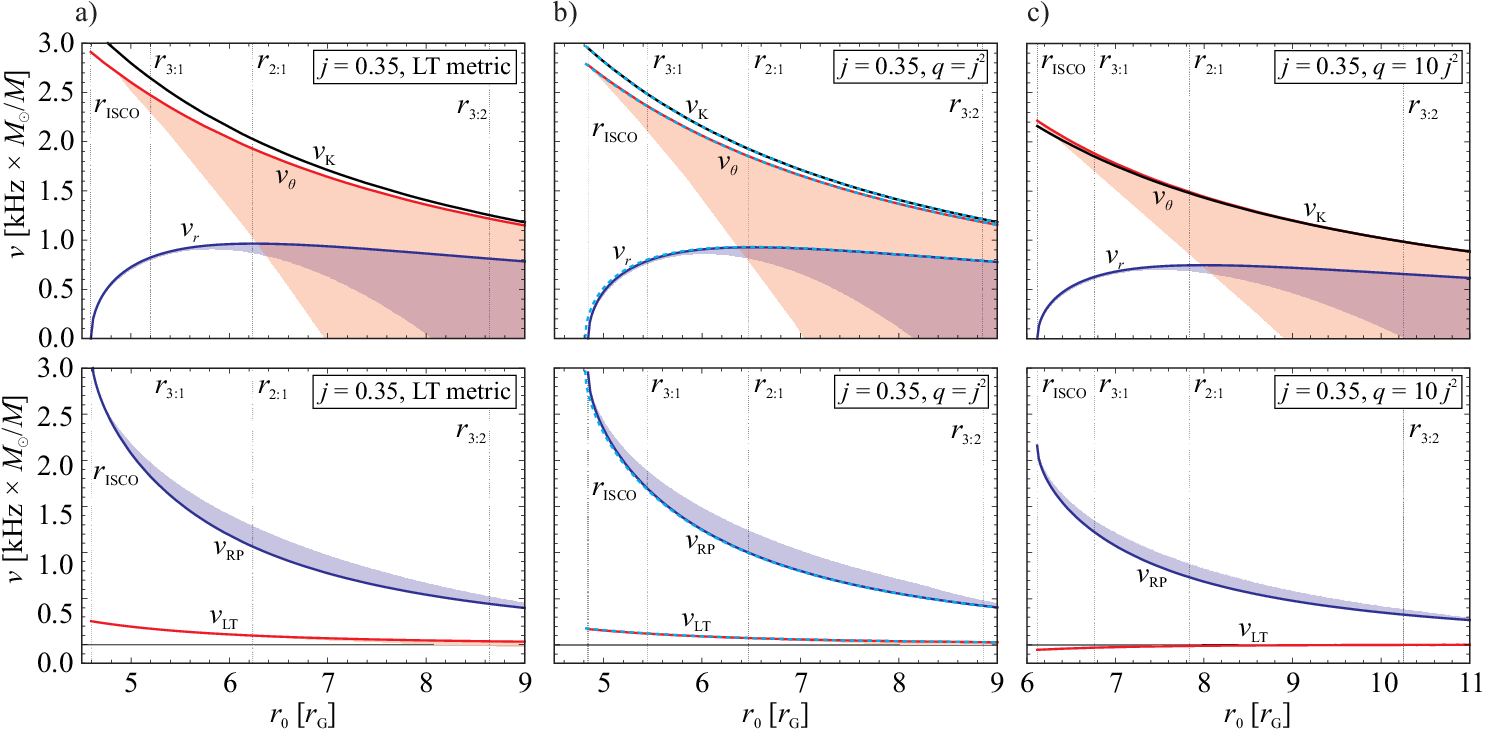}    
   \caption{Frequencies of orbital motion around rotating NSs. a) NS spacetime approximated using the LT metric. b) Spacetime of extremely compact NSs approximated using the Kerr limit of the HT metric. c) Spacetime of highly oblate NSs approximated by the HT metric.  The upper panels show the Keplerian and (axisymmetric) epicyclic frequencies. The bottom panels show the precession frequencies. The full lines correspond to the frequencies of a test particle motion and slender tori. The colour-shaded regions indicate frequencies relevant to the non-geodesic flows of non-slender tori. The dashed lines in panel b indicate the frequencies of a test particle motion in the Kerr spacetime calculated using the appropriate coordinate transformation.}
    \label{fig:freq2}    
\end{figure*}

\begin{figure}
    \centering
    \includegraphics[width=0.88\linewidth]{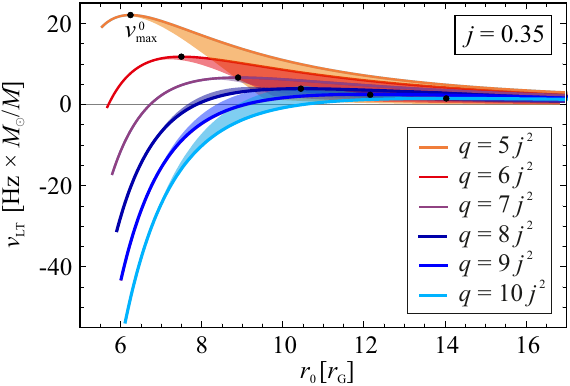}
    \caption{Change in the orientation of the LT precession and behaviour of local frequency maxima shown for $j=0.35$. As in Figure~\ref{fig:freq2}, the solid lines correspond to the frequencies of test particle motion and slender tori, while the colour-shaded regions indicate the frequencies relevant to non-geodesic flows. The full black circles indicate local maxima of the geodesic precession frequency, $\nu_{\mathrm{max}}^0$.}
    \label{figure:LText}
\end{figure}

\begin{figure*}[t]
    \centering
    \includegraphics[width=0.97\linewidth]{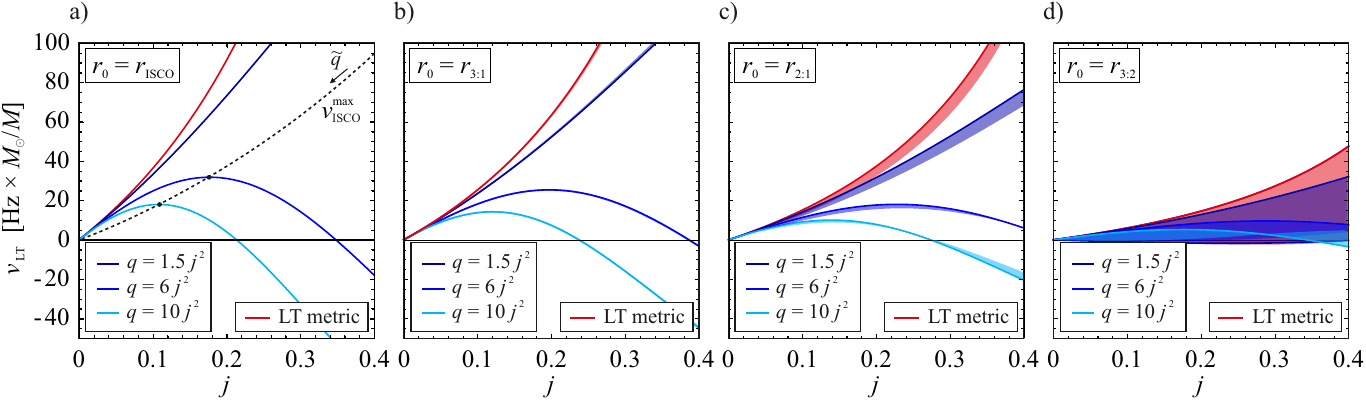}    
    \caption{LT precession frequency in LT and HT spacetimes. a) At the ISCO. b) Close to the ISCO. c) Close to the location of the maximum of the geodesic radial epicyclic frequency. d) At a larger radius, outside of the location of the maximum of the geodesic radial epicyclic frequency. The red colour corresponds to the LT metric. Lines in shades of blue correspond to the HT metric with varying $q$. The displayed full lines show the frequencies relevant to the test particle motion and slender tori. 
    The colour-shaded regions indicate frequencies relevant to non-geodetic flows and cover all allowable torus thicknesses. The dotted line in panel a denotes the maximum frequency at the ISCO given by equation (\ref{equation:jmax}).}   
    \label{fig:lt}
\end{figure*}

\section{LT precession}

In  Schwarzschild geometry, the vertical epicyclic frequency equals the Keplerian frequency, $\Omega_\theta = \OmegaK$. However, in the HT geometry, the vertical epicyclic frequency is different due to the breaking of spherical symmetry and takes the form \citep{abramowicz20003}
\begin{equation}
    \Omega_\theta^2 = \frac{M}{r^3} \left[ 1 - j \delta_1(r) + j^2 \delta_2(r) + q \delta_3(r) \right],
\end{equation}
where the terms $\delta_1(r)$, $\delta_2(r)$, and $\delta_3(r)$ are again provided in the Wolfram Mathematica notebook.

This induces LT precession with a frequency given by
\begin{equation}
\Omega_{\LT}^0 = \OmegaK - \Omega_\theta.
\end{equation}
In the LT spacetime, this precession causes the orbital plane to precess in the direction of the NS's rotation. The profiles of the oscillation frequencies of a free test particle corresponding to the frequencies of axisymmetric oscillations of slender tori are shown in the upper panels of Figure \ref{fig:freq2}, and the bottom panels illustrate the behaviour of the precession frequencies. Within the figure, we assume a moderately rotating star and compare the frequencies calculated within the LT metric (panels a) to those calculated for the Kerr limit of the HT spacetimes corresponding to an extremely compact NS (that nearly coincide with those calculated for the Kerr metric, panels b), and those calculated for a highly oblate NS described in the HT spacetime (panels c).

\subsection{Fluid flows}

To inspect the behaviour of the fluid flow oscillations, we follow \citet{mat2024b,mat2024a} and consider an inner accretion torus orbiting an NS. The adopted torus model assumes a fluid with constant specific angular momentum throughout the flow, with Keplerian value at the torus centre. A polytropic equation of state further describes the fluid. In contrast to the oscillation frequencies of a free test particle, the oscillations of such a fluid torus involve additional corrections due to the internal pressure forces. The oscillation frequencies are given as 
\begin{equation}
    \omega_{i,m} = \omega_{i}^{(0)} + m \OmegaK + \beta^2 \omega_{i,m}^{(2)},
    \label{equation:omega}
\end{equation}
where $\omega_{i}^{(0)}$ represents the epicyclic frequencies of a free test particle, $i = \{r,\theta\}$, $m$ denotes the oscillation mode, parameter $\beta$ defines the torus thickness, and $\omega_{i,m}^{(2)}$ represents the second-order pressure corrections. All the formulae allowing to express these terms explicitly are provided within the associated Wolfram Mathematica notebook.

In the specific case of the lowest-order non-axisymmetric oscillations ($m=-1$), the oscillatory modes of the torus can be seen as torus precession in the corresponding radial or vertical direction, with frequencies\footnote{In the following, we adopt the convention that the negative sign of the frequency corresponds to the precession in the direction opposite to the direction of rotation.}
\begin{equation}
\nu_{\RP} = -\frac{\omega_{r,-1}}{2\pi} ,\quad 
\nu_{\LT}=  -\frac{\omega_{\theta,-1}}{2\pi},
\label{equation:nu}
\end{equation}
which, in the limit of a slender torus ($\beta=0$), are equal to those corresponding to the test particle motion,
\begin{equation}
\nu_{\RP}^{\beta=0} = \nu_{\RP}^0=\frac{\Omega_{\RP}^0}{2\pi}, \quad \nu_{\LT}^{\beta=0} = \nu_{\LT}^0=\frac{\Omega_{\LT}^0}{2\pi}.
\
\end{equation}

The oscillation frequencies of fluid flows of the non-slender tori are shown in the upper panels of Figure \ref{fig:freq2}. Similarly, the frequencies of the periastron and LT precessions of these flows are shown in the lower panels of Figure \ref{fig:freq2}.

\subsection{Inner parts of the flow}

Within Figure \ref{fig:freq2}, we mark four distinct radii. First, there is the position of the innermost stable circular orbit,  $r_{\mathrm{ISCO}}$, where the geodesic radial epicyclic frequency  vanishes,
\begin{equation}
\Omega_r(r_{\mathrm{ISCO}}) = 0.
\label{equation:ISCO}
\end{equation}
Then, there is the radius close to ISCO, $r_{3:1}$,  where $\nu_\theta=3\nu_{r}$, and the radius $r_{2:1}$,  where $\nu_\theta=2\nu_{r}$ and $\nu_r$ reaches a maximum. When the NS spacetime differ from the case of $j=0$, for the maximum of $\nu_r$ there is still $\nu_\theta\approx2\nu_{r}$ \citep[][]{2008AcA....58....1T}. Last, we mark the radius where $2\nu_\theta=3\nu_{r}$. Next, we inspect how $\nu_{\LT}$
changes with the NS spin along these four radii, which cover the innermost parts of an accretion flow.

\section{Effects of spin}

As elaborated in works of \cite{2014PhRvD..89j4001G}, \cite{tsang2016} and \cite{urbancova2019}, for rotating oblate stars, the geodesic vertical epicyclic frequency exceeds the Keplerian frequency for sufficiently high values of $j$ and $q$. In this situation, as illustrated in Figures \ref{fig:freq2}c and \ref{figure:LText}, which are drawn for $j=0.35$, the LT precession changes orientation at a radius close to the ISCO. Its frequency values are then negative below this radius and positive above it, while $\nu_{\mathrm{LT}}$ again vanishes at infinity. Figure~\ref{figure:LText} further illustrates the presence of a local radial maximum of the precession frequency. This maximum may be present even when the LT precession frequency only vanishes at infinity and not at radii close to the ISCO.

Making use of equations (\ref{equation:omega}), (\ref{equation:nu}),(\ref{equation:ISCO}), and the condition 
\begin{equation}
\frac{\partial\nu_{\LT}(j, r=r_\mathrm{ISCO})}{\partial j}=0, 
\end{equation}
we find that, for any given $\tilde q \gtrsim 1.5$, the precession frequency at the ISCO exhibits a maximum in relation to spin, $\nu^\mathrm{max}_\mathrm{ISCO}$, which occurs at
\begin{equation}
    j^\mathrm{max}_\mathrm{ISCO} \doteq \frac{0.95}{\tilde{q}^{\,0.94}}.
    \label{equation:jmax}
\end{equation}
In Figure~\ref{fig:lt}a, we show several examples of the behaviour of $\nu_{\LT}(j, r=r_\mathrm{ISCO})$ for realistic values of $\tilde{q}$. The displayed frequencies are relevant for the inner edge of a thin accretion disc and for a slender torus, which also represents a torus of maximum thickness at this radius. 
Inspecting the Figure, we can see that the precession frequency clearly exhibits a distinct maximum according to relation (\ref{equation:jmax}).

\subsection{Situation above ISCO}

Situations at radii $r_\mathrm{3:1}$, $r_\mathrm{2:1}$, and $r_\mathrm{3:2}$, that are qualitatively similar to $r_\mathrm{ISCO}$, are illustrated within panels b, c, and d of Figure~\ref{fig:lt}. The test particle motion (and slender tori) frequencies are shown using full lines, while the non-slender torus cases are indicated by shaded areas corresponding to tori of all thicknesses allowable according to the results of \cite{mat2024a}. While at $r_\mathrm{3:1}$ almost no quantitative difference exists between the test particle and fluid motion frequencies, at $r_\mathrm{2:1}$, the spread of frequencies given by the freedom in the torus thickness is clearly visible. Nevertheless, the behaviour of the fluid is qualitatively similar to the particle motion. At larger radii, represented by $r_\mathrm{3:2}$, the differences between the individual values of $\tilde{q}$ are smaller but still apparent.
In each panel of Figure~\ref{fig:lt}, we also include the frequencies calculated in the LT spacetime. The frequency values predicted by monotonic functions associated with this spacetime are substantially higher than the maxima associated with the HT spacetime.

\begin{figure*}[t]
    \centering
    \includegraphics[width=0.97\linewidth]{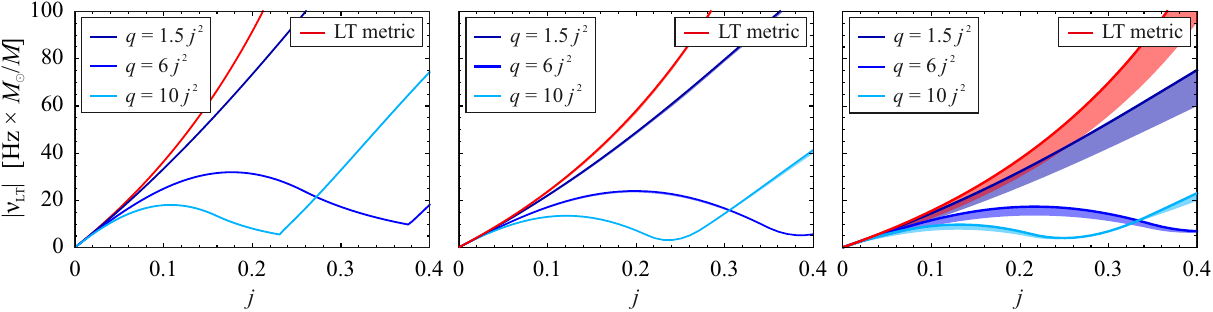}
    \caption{Maximum and average value of the LT frequency. Left panel: Maximum of the frequency, $\nu_{\mathrm{MAX}}$, as a function of the angular momentum of the star. Middle panel: Frequency value averaged from $r_\mathrm{ISCO}$ to $r_\mathrm{2:1}$ as a function of the star's angular momentum. Right panel: Frequency averaged from $r_\mathrm{ISCO}$ to $r_\mathrm{3:2}$ as a function of the star's angular momentum. As in Figure~\ref{fig:lt}, the red lines correspond to the LT metric, while the lines in shades of blue correspond to the HT metric and varying~$q$. The colour-shaded regions then indicate the frequencies relevant to non-geodetic flows and cover the full range of allowable torus thicknesses.}
    \label{fig:ltmax}
\end{figure*}

\subsection{Maximum and averaged frequencies}
\label{section:maximal}

Furthermore, we explore how the maximum possible observable precession frequency, $\nu_{\mathrm{MAX}}$, depends on the star's angular momentum. For low values of $j$ and $\tilde{q}$, it is simply the frequency at the ISCO, $\nu_{\mathrm{MAX}}=\nu_{\mathrm{ISCO}}$.
When there is a a local radial maximum of the geodesic frequency, $\nu_{\mathrm{max}}^0$, which has a higher value than $\lvert \nu_{\mathrm{ISCO}}\rvert$, there is: 
\begin{equation}
\nu_{\mathrm{MAX}} = \nu_{\mathrm{max}}^0,
\end{equation}
otherwise there is
\begin{equation}
\nu_{\mathrm{MAX}} = \lvert \nu_{\mathrm{ISCO}}\rvert,
\end{equation}
which holds overall for any values of $j$ and $q$. As can be seen in Figure~\ref{figure:LText}, the pressure corrections to the geodesic frequency can be positive. In such situations, the maximum frequency associated with the fluid can exceed the value of the maximum geodesic frequency. In this case, however, the maximum absolute value at ISCO is always greater than that for non-geodesic motion. The overall dependence of $\nu_{\mathrm{MAX}}$ on $j$ is shown in the left panel of Figure~\ref{fig:ltmax}.
 
 Finally, similar behaviour is also exhibited by the geodesic frequency averaged along the innermost parts of the disc, which is important when the observed LF QPOs are associated with fluid flows precessing as a rigid body \citep[e.g.][]{ingram2011}. Two examples of such frequencies are illustrated for the radii from ISCO to  $r_{2:1}$ (middle panel of Figure~\ref{fig:ltmax}) and from ISCO to  $r_{3:2}$ (right panel of Figure~\ref{fig:ltmax}). Within these panels, we also indicate the ranges relevant to the frequencies averaged in the same way but for the tori.

\section{Surface}

\begin{figure*}[t]
    \centering
    \includegraphics[width=0.97\linewidth]{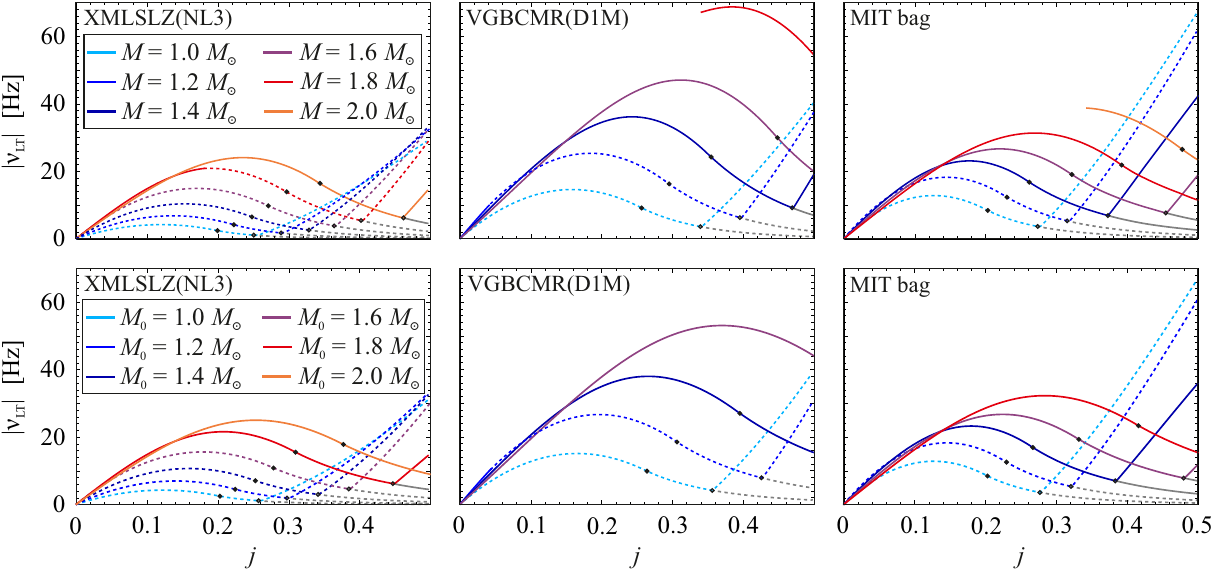}
    \caption{Maximum LT precession frequency as a function of angular momentum of the star based on specific EoS and a fixed gravitational mass, $M$ (top panels), or a fixed non-rotating mass, $M_0$ (bottom panels). Left panels:  XMLSLZ(NL3) EoS. Middle panels: VGBCMR(D1M) EoS. Right panels: MIT bag EoS. The coloured solid lines show the values of $\nu_{\mathrm{MAX}}$ when the star's surface is below the ISCO, while the coloured dashed lines show the values of $\nu_{\mathrm{MAX}}$ when the surface is above the ISCO. The circles on the individual lines enable us to determine the radius at which the maximum displayed frequency is reached. For values of $j$ below that indicated by the black filled circle (or when there is no circle), the frequency corresponds to the edge of the relevant radial interval: the ISCO (solid lines) or the NS surface (dotted lines). For higher values of $j$, below the value indicated by the open circle, the frequency corresponds to the maximum above the edge. Above this value, it again corresponds to the ISCO for the solid lines and the NS surface for the dotted lines. The grey lines then indicate the value of the corresponding local geodesic maximum, $\nu_{\mathrm{max}}^0$.}
    \label{fig:ltmaxeos}
\end{figure*}

The investigation above does not take into account the restrictions imposed by the location of the compact object's surface \citep[e.g.,][]{2014A&A...564L...5T,2015MNRAS.454.4066P,urbancova2019}. To illustrate these restrictions, we consider three different specific EoSs. The first two are chosen from the public CompOSE EoS library \citep{Compose} in such a way that they represent opposite margins of the overall region spanned by all COMPOSE EoSs in the mass-radius diagram. These are XMLSLZ \citep[NL3;][]{XMLSLZ} and VGBCMR \citep[D1M;][]{VGBCMR} EoSs, respectively. The third is the MIT~bag EoS \citep{1974PhRvD..10.2599C}, representing QSs, whereas we used the standard values $B=10^{14} \,\mathrm{g\,cm}^{-3}$ for the bag constant and $\alpha_\mathrm{c} = 0.15$ for strong interaction coupling constant.

Within the top panels of Figure~\ref{fig:ltmaxeos}, we use the three EoS chosen above and plot the maximum LT precession frequency as a function of the NS angular momentum. The solid lines show the values when the star's surface is below the ISCO, and the dashed lines show the values when it is above the ISCO.
Furthermore, the plots allow us to determine whether the maximum frequency is reached at the edge of the relevant radial interval — the ISCO or NS surface — or above it. For practical purposes, it can also be useful to consider a sequence of stars with a constant non-rotating mass, $M_0$, rather than a constant gravitational mass, $M$. This is illustrated in the bottom panels of Figure \ref{fig:ltmaxeos}, which show a full analogy of the top panels, except that the fixed quantity is $M_0$, while $M$ increases from $M=M_0$ towards higher values as $j$ increases from $j=0$.

Examining the Figure \ref{fig:ltmaxeos} shows that, although stars with low masses extend above the ISCO, the effect of spin on the achieved value of the LT precession frequency is still similar for stars with high and low masses.

\section{Discussion and conclusions}
\label{sec:discussion}

The works of \cite{2012MNRAS.422.2581P}, \cite{kluzniak2013,kluzniak2014}, \cite{2014PhRvD..89j4001G}, \cite{2015MNRAS.454.4066P}, \cite{tsang2016}, and \cite{mat2024b,mat2024a} have revealed the importance of the interplay between relativistic effects and effects already present in the Newtonian physics of oblate bodies. In the case of LT precession, this interplay takes place between the contributions of relativistic and classical precession, which have opposite directions, and this is crucial for our study utilising the HT spacetimes.
 
 In contrast to the LT spacetime, which is linear in NS angular momentum, in the HT spacetime, the dependence of the frequency on $j$ is not monotonic and exhibits maxima even for a moderately oblate NS. Inspecting Figures  \ref{fig:lt}--\ref{fig:ltmaxeos}, we conclude that, within the innermost accretion region, the contribution of the relativistic precession to the overall precession ensures the growth of the total precession frequency only for a small interval of $j$. Afterwards, the frequency decreases due to the increasing contribution of the classical precession. Then, for higher $j$, the contribution of the classical precession drives the increase of the absolute frequency value, while the orientation of the resulting precession is reversed.

Consequently, within a sequence of NSs of equal mass and increasing rotational frequency, reaching moderate values of NS angular momentum, the precession frequencies can exhibit a sharp maximum at a particular value of the star's rotational frequency. This conclusion, which holds for both geodesic and non-geodesic accretion flows, is relevant to the models of NS LF QPOs and implies that very different groups of accreting NSs, slow and fast rotators, can exhibit the same observable precession frequencies.\footnote{Taking into account the thickness of the flow effectively implies a small quantitative correction to these frequencies when considering the innermost accretion region within a few gravitational radii.} This may partially explain the previous observations, which revealed the lack of evidence for the expected correlation between the frequencies of QPOs and NS spin.

\begin{acknowledgements}
 We thank to the referee for valuable comments and suggestions. We acknowledge the Czech Science Foundation (GA\v{C}R) grants no. 21-06825X and 25-16928O, the MSK grant no. 04076/2024/RRC and the internal grant of the Silesian University in Opava no. $\mathrm{SGS/25/2024}$. 
\end{acknowledgements}

\bibliographystyle{aa}
\bibliography{LTprecession}

\end{document}